\documentclass[twocolumn,english,letterpaper,superscriptaddress,longbibliography,prl]{revtex4-1}

\usepackage[T1]{fontenc}
\usepackage[latin9]{inputenc}
\usepackage{color}
\usepackage{amsmath}
\usepackage{amssymb}
\usepackage{graphicx}
\usepackage{esint}
\usepackage{babel}
\begin{document}

\title{Equilibration Dynamics of Strongly Interacting Bosons in 2D Lattices
with Disorder}

\author{Mi Yan}

\affiliation{Department of Physics, Virginia Tech, Blacksburg, VA 24061, USA}

\author{Hoi-Yin Hui}

\affiliation{Department of Physics, Virginia Tech, Blacksburg, VA 24061, USA}

\author{Marcos Rigol}

\affiliation{Department of Physics, The Pennsylvania State University, University
Park, PA 16802, USA}

\author{V.W. Scarola}

\affiliation{Department of Physics, Virginia Tech, Blacksburg, VA 24061, USA}
\begin{abstract}
Motivated by recent optical lattice experiments {[}J.-y. Choi et al.,
Science \textbf{352}, 1547 (2016){]}, we study the dynamics of strongly
interacting bosons in the presence of disorder in two dimensions.
We show that Gutzwiller mean-field theory (GMFT) captures the main
experimental observations, which are a result of the competition between
disorder and interactions. Our findings highlight the difficulty in
distinguishing glassy dynamics, which can be captured by GMFT, and
many-body localization, which cannot be captured by GMFT, and indicate
the need for further experimental studies of this system.
\end{abstract}

\date{\today}

\maketitle
\emph{Introduction}.\textemdash Ultracold atoms loaded into optical
lattices \cite{verkerk:1992,jessen:1992,hemmerich:1993} offer ideal
platforms to study localization \cite{bloch:2008,cazalilla:2011}.
Examples in the noninteracting limit include fermionic band insulators
\cite{kohl:2005}, and, in the presence of (quasi-)disorder, Anderson
insulators \cite{billy:2008a,roati:2008,chabe:2008,lemarie:2009,kondov:2011,jendrzejewski:2012}.
In clean systems, localization can also occur because of interactions,
producing Mott insulators (MIs) \cite{greiner:2002,stoferle:2004,spielman:2007,jordens:2008,schneider:2008}.
Recent experimental studies have explored the interplay between disorder
and interactions \cite{chen:2008,white:2009,pasienski:2010,gadway:2011,beeler:2012,brantut:2012,tanzi:2013,krinner:2013,kondov:2015,schreiber:2015,choi:2016}.
In the ground state of bosonic systems, this interplay can generate
the Bose-glass (BG) phase \cite{fisher:1989,scalettar:1991}. The
BG, like the bosonic MI, is characterized by a vanishing superfluid
density but, unlike the MI, it is compressible. At extensive energy
densities above the ground state, the interplay between disorder and
interactions can lead to a remarkable phenomenon known as many-body
localization (MBL) \cite{basko:2006,oganesyan:2007,pal:2010}. In
the MBL phase, eigenstate thermalization \cite{deutsch:1991,srednicki:1994,rigol:2008}
does not occur \cite{nandkishore:2015}.

Signatures of MBL were recently observed with fermions \cite{kondov:2015,schreiber:2015}
and bosons in two dimensions (2D) \cite{choi:2016}. Our work is motivated
by the latter experiment (see Refs.~\cite{scarola:2015,mondaini:2015}
for theoretical studies inspired by the former). In Ref.~\cite{choi:2016},
a MI with one boson per site was prepared in a harmonic trap in a
deep optical lattice. All bosons in one half of the system were then
removed and the remaining half was allowed to evolve by lowering the
lattice depth, with or without disorder. During the dynamics, the
parity-projected occupation of the lattice sites was measured using
fluorescence imaging, allowing the study of the evolution of the imbalance
${\cal I}$ between the initially occupied and unoccupied halves.
With no or weak disorder, ${\cal I}$ vanished within times accessible
experimentally, i.e., it attained the value expected in thermal equilibrium.
But beyond a certain disorder strength, ${\cal I}$ appeared to saturate
to a nonzero value. This saturation was taken as evidence for MBL
\cite{choi:2016}.

Features of the experimental setup in Ref.~\cite{choi:2016} can
lead to a very slow equilibration of ${\cal I}$ to the point of making
it difficult to distinguish glassy behavior from the MBL phase. First,
the initial dynamics in the unoccupied half of the trap is dominated
by Anderson physics (because of low site occupations). Second, the
initial MI, before the removal of the bosons in one half of the system,
is close in energy to the ground state after the lattice depth is
lowered but the system remains deep in the MI regime. The latter MI,
in turn, is close in energy to a BG with a site occupancy slightly
below one at the same interaction strength (if the disorder is strong
enough to generate a BG). Therefore, the dynamics resulting from the
gradual decrease of the site occupations in the occupied half of the
system, after the removal of the bosons in the other half, can be
dominated by excitations of the BG in the remaining half.

To study the impact of glassy physics we use Gutzwiller mean-field
theory (GMFT) to model the dynamics of the experiments in Ref.~\cite{choi:2016}.
GMFT provides qualitatively correct phase diagrams for strongly interacting
clean \cite{rokhsar:1991,jaksch:1998,hen:2009,hen:2010} and disordered
\cite{sheshadri:1995,damski:2003,buonsante:2007a,buonsante:2009,lin:2012}
(away from the tip of the Mott lobe) systems. It has also been used
to study non-equilibrium effects such as the dynamical generation
of molecular condensates \cite{jaksch:2002} and MIs \cite{zakrzewski:2005},
dipole oscillations \cite{snoek:2007}, quantum quenches \cite{wolf:2010,snoek:2011,lin:2012},
expansion dynamics \cite{hen:2010a,jreissaty:2011}, and transport
in the presence of disorder \cite{lin:2012,yan:2016}. However, since
the Gutzwiller ansatz wavefunction is a product state, it has zero
entanglement entropy for any partitioning of the system. GMFT is therefore
capable of capturing BG dynamics but it cannot capture thermalization
and MBL phases \cite{bauer:2013}, which after taken out of equilibrium,
e.g., using a quantum quench, exhibit a linear \cite{kim:2013} and
logarithmic \cite{bardarson:2012} growth of the entanglement entropy,
respectively, with time.

We use GMFT to study the dynamics of initial states under the same
(or similar) conditions as the experiment, thus allowing direct comparison.
We find that the GMFT dynamics is similar but not quite the same as
that in the experiment. In particular, the GMFT state rebalances more
slowly, which motivates us to add a phenomenological parameter to
our theory to gradually remove slow particles from data analysis because
their dynamics are not accurately captured by our theory. A single
phenomenological parameter significantly improves the agreement between
theory and experiment.

Given the fact that GMFT cannot describe dynamics in a MBL phase,
our results raise concerns as to whether experimental observations
are the result of MBL or the result of slow transport due to glassy
dynamics. Only the latter is captured by our GMFT treatment.

\emph{Model.\textemdash }We consider bosons in a 2D square lattice
subjected to disorder and a parabolic trapping potential, as described
by the Bose-Hubbard Hamiltonian,
\begin{equation}
\hat{H}=-J\sum_{\left\langle {\bf {\bf i}j}\right\rangle }\hat{b}_{{\bf i}}^{\dagger}\hat{b}_{{\bf {\bf j}}}^{\vphantom{\dagger}}+\frac{U}{2}\sum_{{\bf {\bf i}}}\hat{n}_{{\bf i}}\left(\hat{n}_{{\bf i}}-1\right)+\sum_{{\bf i}}\mu_{{\bf i}}\hat{n}_{{\bf i}},\label{eq:H}
\end{equation}
where $\hat{b}_{{\bf {\bf i}}}^{\dagger}$ creates a boson at site
${\bf i}\equiv(i_{x},i_{y})$ and $\hat{n}_{{\bf {\bf i}}}=\hat{b}_{{\bf i}}^{\dagger}\hat{b}_{{\bf i}}$
is the site occupation operator. $J$ parametrizes the tunneling between
nearest neighbors and $U$ is the on-site repulsive interaction. The
chemical potential ($\mu$), harmonic trap (of strength $\Omega$),
and disorder potential ($\epsilon_{{\bf i}}$) are in $\mu_{{\bf i}}=-\mu+\Omega\left|{\bf i}-{\bf r}_{0}\right|^{2}+\epsilon_{{\bf i}}$,
with ${\bf r}_{0}=(0,0)$. We focus on a lattice with $31\times31$
sites in which, for the Hamiltonian parameters used here, the sites
at the edges are always empty. We consider two types of disorder,
with uniform and Gaussian distributions, whose strengths are denoted
by $\Delta_{u}$ and $\Delta_{g}$, respectively. We set $k_{B}=\hbar=1$.

\emph{Methods}.\textemdash We study the dynamics of zero and nonzero
temperature initial states. The density matrix within GMFT is
\begin{equation}
\hat{\rho}\left(t\right)=\prod_{{\bf i}}\hat{\rho}_{{\bf i}}\left(t\right)=\prod_{{\bf i}}\left[\sum_{m,n=0}^{\infty}\alpha_{mn}^{\left({\bf i}\right)}\left(t\right)\left|m\right\rangle _{{\bf i}{\bf i}}\left\langle n\right|\right],\label{eq:GutzDM}
\end{equation}
where $\left|n\right\rangle _{{\bf i}}$ is the state with $n$ bosons
at site ${\bf {i}}$, and $t$ denotes time. This ansatz decouples
Eq.~\eqref{eq:H} into single-site Hamiltonians $\hat{H}_{{\bf i}}^{{\rm MF}}=-J(\phi_{{\bf i}}^{*}\hat{b}_{{\bf {\bf i}}}^{\vphantom{\dagger}}+\phi_{{\bf i}}\hat{b}_{{\bf {\bf i}}}^{\dagger})+(U/2)\sum_{{\bf i}}\hat{n}_{{\bf i}}(\hat{n}_{{\bf i}}-1)+\mu_{{\bf i}}\hat{n}_{{\bf i}}$,
where $\phi_{{\bf i}}=\sum_{{\bf {\bf j}\in{\rm nn}_{{\bf i}}}}{\rm Tr}(\hat{\rho}_{{\bf j}}\hat{b}_{{\bf j}})$
sums over neighbor sites to ${\bf {\bf i}}$. Substituting Eq.~\eqref{eq:GutzDM}
into the von Neumann equation, $i\partial_{t}\hat{\rho}=[\hat{H},\hat{\rho}]$,
leads to the equation of motion for $\alpha_{mn}^{({\bf {i})}}$:
\begin{align}
i\partial_{t}\alpha_{m,n}^{\left({\bf i}\right)}= & -J\phi_{{\bf i}}^{*}\left[\sqrt{m+1}\alpha_{m+1,n}^{\left({\bf i}\right)}-\sqrt{n}\alpha_{m,n-1}^{\left({\bf {\bf i}}\right)}\right]\nonumber \\
 & -J\phi_{{\bf i}}\left[\sqrt{m}\alpha_{m-1,n}^{\left({\bf i}\right)}-\sqrt{n+1}\alpha_{m,n+1}^{\left({\bf i}\right)}\right]\nonumber \\
 & +\frac{U}{2}\left[m\left(m-1\right)-n\left(n-1\right)\right]\alpha_{m,n}^{\left({\bf i}\right)}\nonumber \\
 & +\mu_{{\bf i}}\left(m-n\right)\alpha_{m,n}^{\left({\bf i}\right)},\label{eq:eqmot}
\end{align}
which yields the time evolution of the site occupations: $n_{{\bf {\bf i}}}(t)={\rm Tr}(\hat{\rho}_{{\bf i}}\hat{n}_{{\bf i}})$.

Following Ref.~\cite{choi:2016}, we quantify the degree of localization
using the imbalance, 
\begin{equation}
{\cal I}(t)=\frac{N_{L}(t)-N_{R}(t)}{N_{L}(t)+N_{R}(t)},\label{eq:imbalance}
\end{equation}
where $N_{L}(t)=\sum_{-l_{x}\leq i_{x}<i_{0},|i_{y}|\leq l_{y}}n_{i_{x},i_{y}}(t)$
and $N_{R}(t)=\sum_{i_{0}\leq i_{x}\leq l_{x},|i_{y}|\leq l_{y}}n_{i_{x},i_{y}}(t)$,
with an $l_{x}\times l_{y}$ central region of interest. $l_{y}$
is taken to be 2 to define a window five lattice sites wide in the
$y$ direction. We first set $l_{x}$ to $l_{W}=9$, as in experiment.
In Ref.~\cite{choi:2016}, the lattice center does not always coincide
with the center of the harmonic potential, and this causes an imperfect
preparation of the initial state domain wall. To account for this,
the line separating the left and right sides of the system is defined
using $i_{0}=0$ or $i_{0}=1$. The imbalance is obtained by averaging
the two cases.

We also compute the inverse decay length $\lambda(t)$ \cite{choi:2016}.
To calculate $\lambda(t)$ we first compute the average, $\bar{n}_{i_{x}}(t)=(2l_{y}+1)^{-1}\sum_{|i_{y}|\leq l_{y}}n_{i_{x},i_{y}}(t)$.
$\lambda$ is then obtained by fitting
\begin{equation}
\bar{n}_{i_{x}}\left(t\right)/\bar{n}_{i_{x}}^{0}\sim e^{-\lambda\left(t\right)\,i_{x}},\label{eq:decayfit}
\end{equation}
where $\bar{n}_{i_{x}}^{0}$ is the zero disorder steady-state density
and $i_{x}$ denotes a least squares fit from $i_{x}=0$ to $l_{x}$.

For $\hat{\rho}\left(t=0\right)$, we take the ground state or a thermal
state of the initial Hamiltonian, such that $\hat{\rho}_{{\bf {\bf i}}}=Z_{{\bf i}}^{-1}e^{-\beta\hat{H}_{{\bf i}}^{{\rm MF}}}$
(where $\beta=1/T$ is the inverse temperature and $Z_{{\bf i}}$
is the partition function). Our calculations in the presence of disorder
are done for an ensemble of disorder realizations. Disorder averaging
over around 100 disorder realizations is sufficient for convergence.

\begin{figure}
\begin{centering}
\includegraphics[width=0.7\columnwidth]{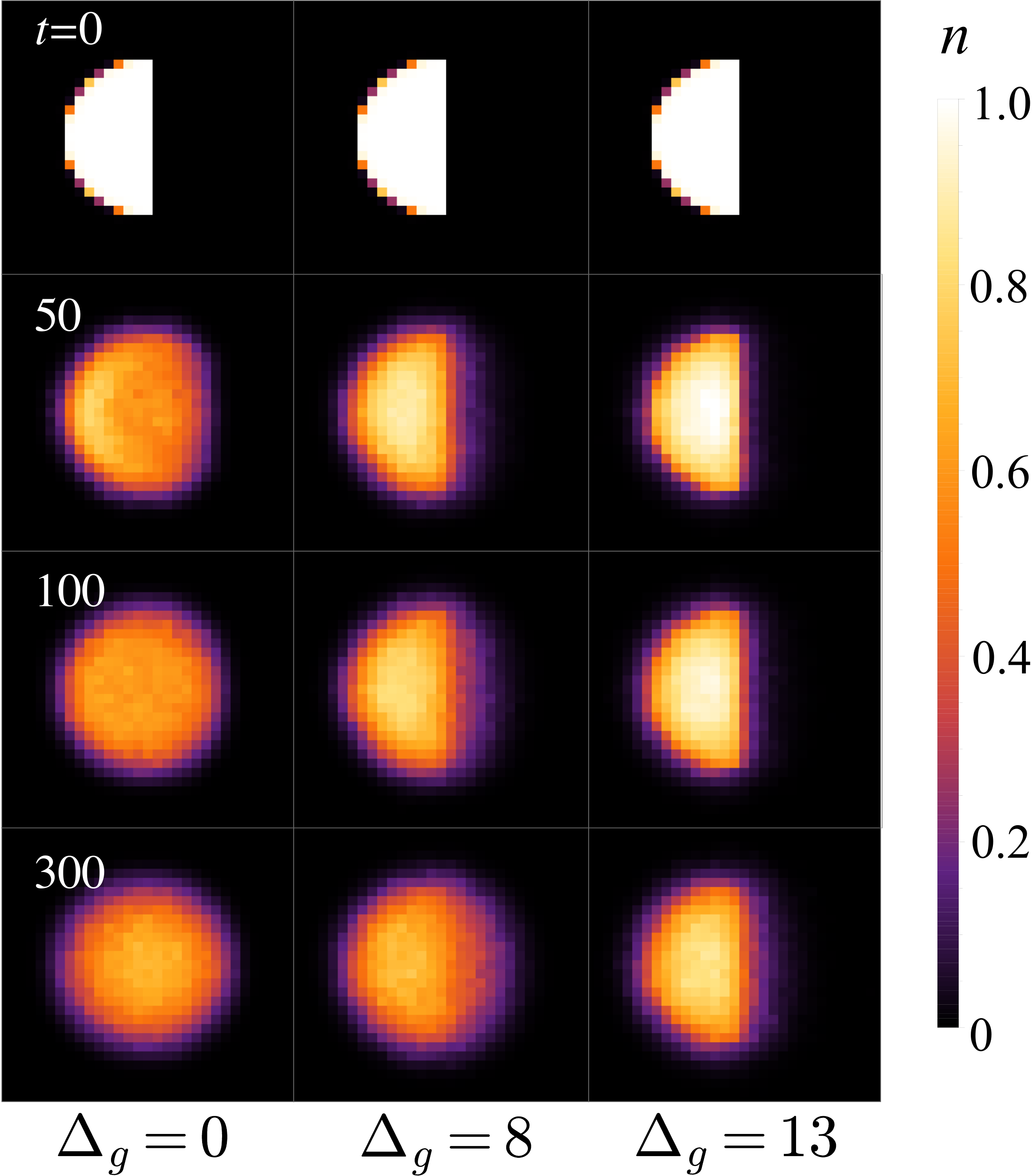} 
\par\end{centering}
\caption{\label{fig:density}The site occupations for quenched dynamics at
zero temperature. Columns (rows) depict results for different disorder
strengths (different times). At time $t=0$ all bosons in the right
half of the system are removed and the remainder evolves for $t>0$.
The $t=0$ state is the ground state for a very small hopping and
no disorder. For $t>0$, Gaussian disorder of strength $\Delta_{g}$
is introduced and the hopping is increased. The state evolves for
$t\geq0$ with no parameter changes.}
\end{figure}

Within GMFT, dynamics occur only when there are nonvanishing values
of the order parameter $\phi_{{\bf {i}}}$ {[}see Eq.~\eqref{eq:eqmot}{]}.
As a result, a pure MI state would exhibit no dynamics within GMFT.
We find that, as in Refs.~\cite{jreissaty:2011,yan:2016}, the small
region with a non-vanishing order parameter generated by the harmonic
trap at the edge of MI domains is sufficient to drive dynamics. Remarkably,
we will see that the ensuing dynamics measured by imbalance is slower
but similar to that in the experiments \cite{choi:2016} at long times.
We will then show that decreasing $l_{x}$ to phenomenologically remove
particles in the MI state significantly improves agreement with experiment.

\begin{figure}
\begin{centering}
\includegraphics[width=0.9\columnwidth]{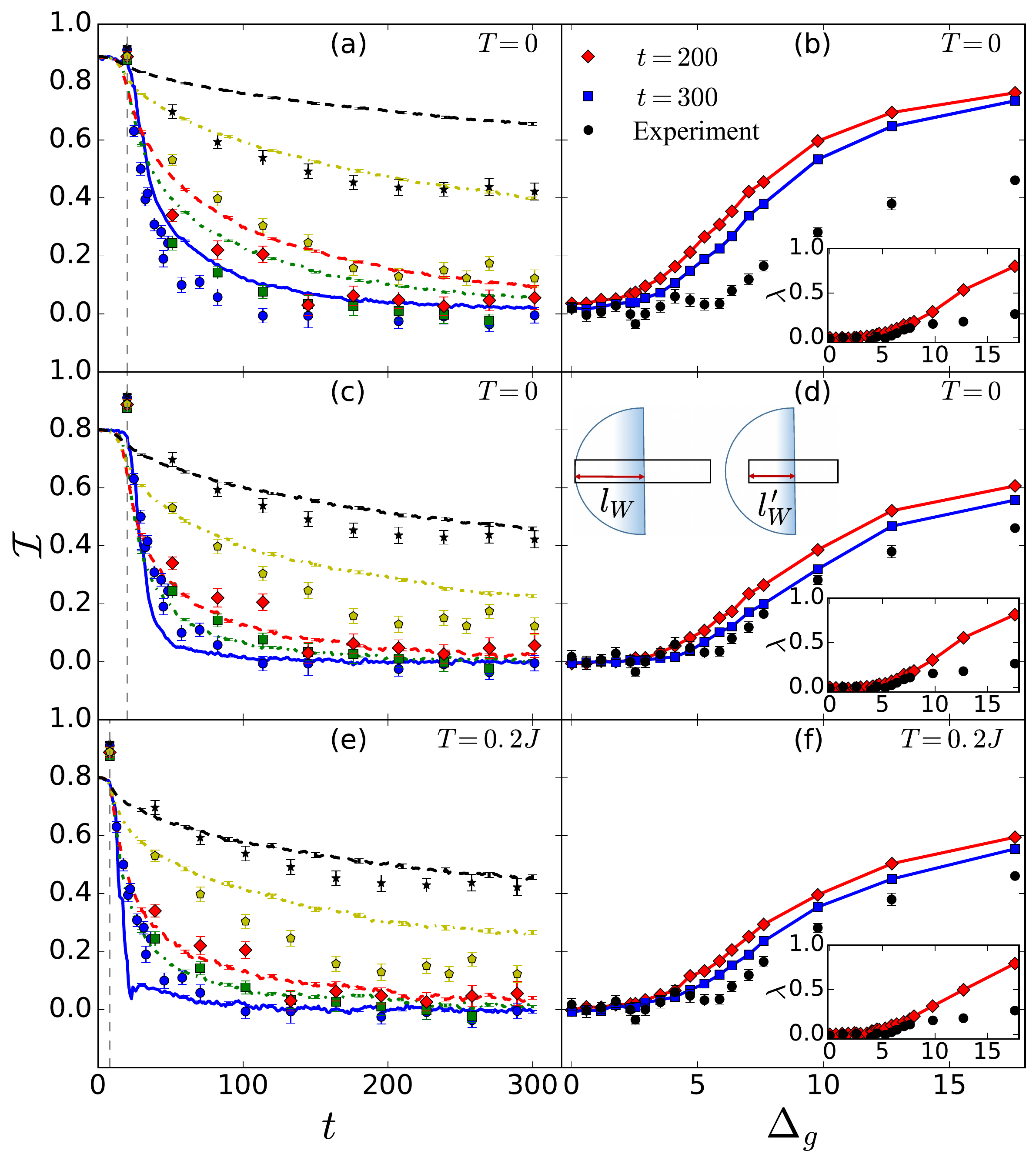} 
\par\end{centering}
\caption{\label{fig:quenched}(a): Time evolution of the imbalance ${\cal I}$
for various disorder strengths at initial temperature $T=0$. Lines
show simulation results while points with error bars show experimental
data \cite{choi:2016}. The vertical dashed line marks a time $t^{*}=20$
below which ${\cal I}$ barely changes within GMFT. The experimental
results are shifted to start at $t^{*}$. From bottom to top the lines
and symbols correspond to $\Delta_{g}=0,3,4,8,$ and $13$. (b): Corresponding
${\cal I}$ for the same parameters but at times $t=200$ and $300$
against disorder strength. The experimental result after an evolution
time of $187$ are also plotted. The inset shows the inverse decay
length {[}Eq.~(\ref{eq:decayfit}){]} from our calculation at $t=200$
and for the experiment after an evolution time $t=187$. (c) and (d):
The same as (a) and (b) but with an analysis window resized from $l_{W}=9$
{[}as in (a) and (b) following Ref.~\cite{choi:2016}{]} to $l_{W}^{\prime}=5$,
as shown in the schematic. (e) and (f): The same as (c) and (d) but
at non-zero temperature. Here $t^{*}$ reduces to 8.}
\end{figure}

\emph{Qenched dynamics.\textemdash{}} In the experiment \cite{choi:2016}
the dynamics took place after lowering the lattice depth and introducing
a disorder potential to a MI created in a deep lattice and to which
all atoms in one half of the system were removed. 
From now on, we use the hopping parameter after the quench $J=U/24.4$
as our energy unit. To create the initial state, we used the experimental
parameters \cite{choi:2016}: $J_{I}=0.244$, $U=24.4$, $\Omega=0.145$,
and $\mu=10.6$. After free energy minimization, particles on the
right half of the system are manually removed {[}by setting $\alpha_{m,n}^{(i_{x}>0)}=\delta_{m,0}\delta_{n,0}$
in Eq.~\eqref{eq:GutzDM}{]}, leaving behind a particle number comparable
with the experiment, $\ensuremath{N_{b}\approx123}$. In accordance
with the experimental protocol \cite{choi:2016}, to generate disorder
(at the evolution stage) we square a two-dimensional array of uniformly
distributed random numbers followed by a convolution with a Gaussian
profile of standard deviation $0.5$. The disorder strength $\Delta_{g}$
is defined as the full width at half maximum of the resulting disorder
profile.

The first column in Fig.~\ref{fig:density} depicts the evolution
of the site occupations in the absence of disorder. Here the particles
expand to reach a steady state with no imbalance. When disorder of
strength $\Delta_{g}=8$ is introduced, the motion slows considerably
and an imbalance remains at the latest time shown. For very strong
disorder ($\Delta_{g}=13$), the particles remain almost entirely
in the initially occupied region.

To quantitatively understand the dynamics, we plot the imbalance against
time in Fig.~\ref{fig:quenched}(a). For $t<t^{*}$, the imbalance
barely changes. This is an artifact of GMFT for the initial state,
which is mostly a MI domain. Beyond $t^{*}$, ${\cal I}$ vanishes
rapidly in the clean limit and for weak disorder. But, as the disorder
strength increases, it takes longer for ${\cal I}$ to reach the expected
${\cal I}=0$ steady state value. In Fig.~\ref{fig:quenched}(a),
we also plot the experimental results taking $t^{*}$ to be the starting
time for the experiments. The GMFT and experimental results exhibit
good agreement for weak disorder strength, but the latter exhibit
faster relaxation as the disorder strength is increased.

In Fig.~\ref{fig:quenched}(b), we plot the imbalance alongside experimental
results \cite{choi:2016}, as a function of the disorder strength.
In our theoretical results, the upturn in ${\cal I}$ versus $\Delta_{g}$
moves toward stronger disorder strengths as $t$ increases. A similar
trend was seen in experiments for $t\lesssim200$, but the experimental
results appeared to saturate for $200\lesssim t\lesssim300$. For
any given selected time, the upturn in ${\cal I}$ versus $\Delta_{g}$
occurs at a smaller value of $\Delta_{g}$ in GMFT when compared to
the experiments, which is expected given the slower dynamics of the
former seen in Fig.~\ref{fig:quenched}(a).

$\lambda$ offers another way to quantify the degree of localization
by parameterizing the extent to which disorder suppresses the relaxation
of site occupations. The inset in Fig.~\ref{fig:quenched}(b) shows
$\lambda$ versus $\Delta_{g}$ for $t=200$ and the experimental
results for $t=187$. The behavior of $\lambda$ (inset) is similar
to that of ${\cal I}$ (main panel).

There are also differences between GMFT and experiments. For example,
at weak disorder strengths, the experimental data of Fig.~\ref{fig:quenched}(b)
exhibits oscillations not captured by GMFT. These oscillations in
turn impact the comparison of the nature of upturns of ${\cal I}$
or $\lambda$ near $\Delta_{g}=5.5$, as they make it look sharper
in the experimental results.

A key observable in identifying localization is the time derivative
of the imbalance, $\dot{{\cal I}}$, at long times, as used in observations
of Anderson localization with ultracold atoms \cite{billy:2008a,roati:2008,chabe:2008,lemarie:2009,kondov:2011,jendrzejewski:2012}.
The vanishing of $\dot{{\cal I}}$ at long times (and in large system
sizes) is a necessary condition for localization. The slope of ${\cal I}$
versus $t$ obtained for the four latest experimental times reported
is $-1.017(\pm1.028)\times10^{-4}$ for the largest disorder strength.
Here we see that the experimental error is too large to definitively
show a vanishing of the slope since the results are also consistent
with just a small slope. Within GMFT, we find a small non-zero slope:
$-4.433(\pm0.053)\times10^{-4}$, for the largest disorder strength.
The small non-zero slope shows that a slow rebalancing (as expected
in the glassy state captured within GMFT) is consistent with experiment.

\begin{figure}
\begin{centering}
\includegraphics[width=0.9\columnwidth]{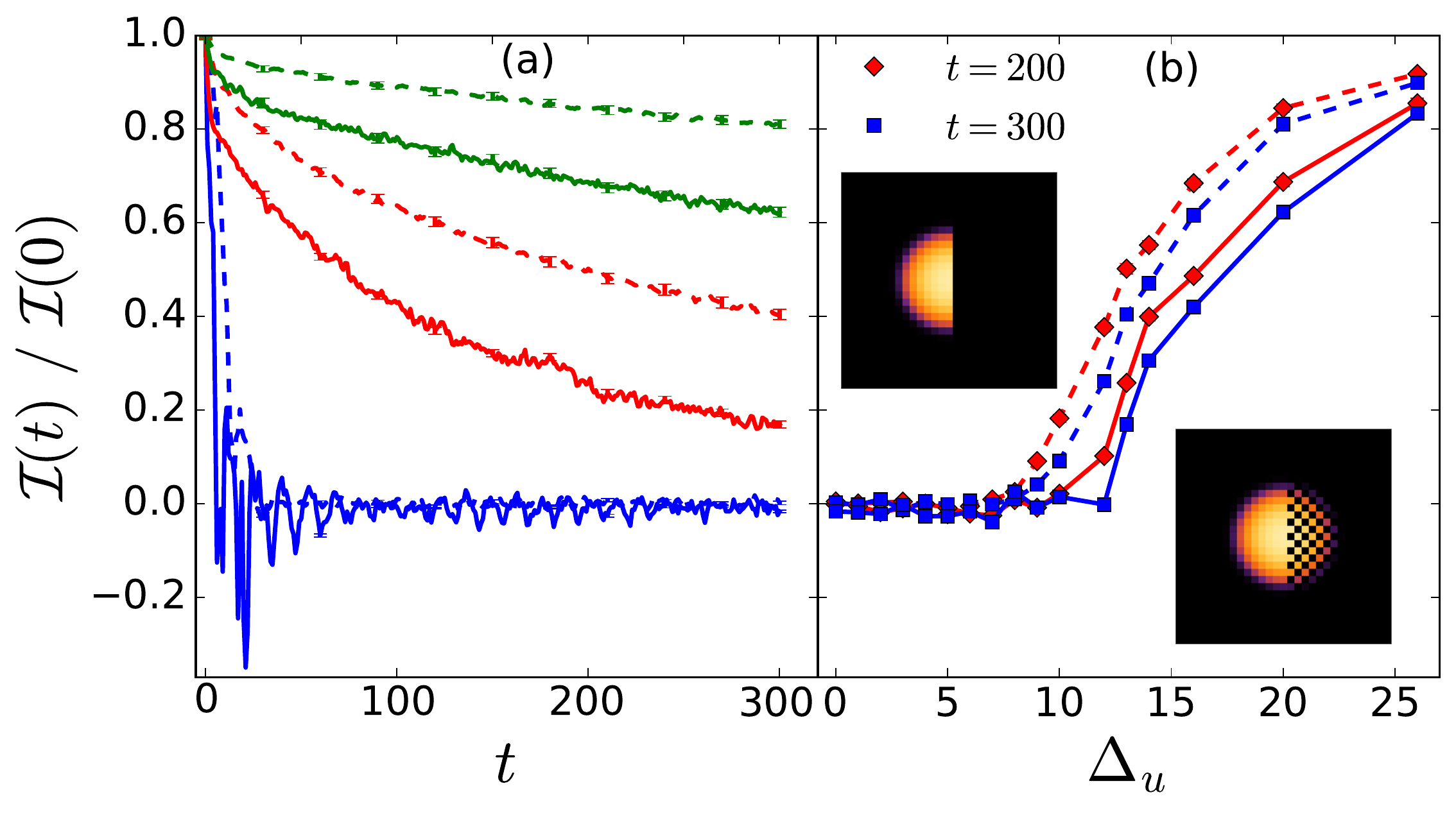} 
\par\end{centering}
\caption{\label{fig:checkerboard} The solid (dashed) lines plot the normalized
imbalance where the right half of the trap was initialized to a checkerboard
(empty) pattern as shown in the insets. (a) The normalized imbalance
against time for various disorder strengths. The pairs of (solid/dashed)
lines correspond to uniform disorder $\Delta_{u}=0,13$ and $20$
from bottom to top. (b) The normalized imbalance at times $t=200$
and $300$ against disorder strength. The other parameters are $U=24.4$,
$\Omega=0.145$, and $\mu=4$.}
\end{figure}

To understand the robustness of our findings within GMFT, we have
also studied initial states at finite temperatures, different quench
protocols, and dynamics in the presence of a uniform disorder distribution.
The appendix shows that the latter two changes do not have much impact
on the imbalance dynamics at long times. GMFT shows that the imbalance
dynamics of a BG or MI quenched into a disorder profile respond in
nearly the same way.

\emph{Phenomenological parameter}.\textemdash To improve the comparison
with the experiments we introduce a phenomenological parameter that
excludes particles which move too slowly within GMFT. GMFT underestimates
the speed of the MI dynamics under an applied field. The motion of
the entire trapped system is therefore slower in GMFT at long times.

To account for the slow Mott particles we introduce a phenomenological
parameter to our GMFT analysis. The inset of Fig.~\ref{fig:quenched}(d)
shows a schematic of a resizing of the window used to compute the
imbalance. The rectangles in the schematics indicate a decrease in
$l_{x}$ in Eq.~\ref{eq:imbalance}, from $l_{W}$ to $l_{W}^{\prime}$.
Our phenomenological parameter, $l_{x}$, therefore increases the
relative rate of rebalancing because slow moving Mott particles near
the left edge of the system are excluded from the data analysis. Decreasing
$\mu$ also removes these particles. We find that tuning either $\mu$
or $l_{x}$ allows us to fit ${\cal I}$ versus $t$ to experimental
values with the same accuracy. We choose $l_{x}$ as our phenomenological
parameter and vary it to obtain a best fit for the largest disorder,
$\Delta_{g}=13$.

Figures~\ref{fig:quenched}(c) and \ref{fig:quenched}(d) plot the
same as panels (a) and (b) but with the new window size, $l_{W}^{\prime}$.
Here there is much better agreement with experiment because the relative
fraction of mobile to localized particles in our GMFT is closer to
the experiment. Panels (e) and (f) include nonzero temperature. In
varying $T$ we find little change for $T<J$. $T=0.2J$ was chosen
as a best fit for the largest disorder. In Fig.~\ref{fig:quenched}(e)
we see that $t^{*}$ diminishes and the imbalance tends to level off
quicker at long times, with a slight increase in the slope to $-4.816(\pm0.160)\times10^{-4}$.

The comparisons between theory and experiment in Fig.~\ref{fig:quenched}
show that by adjusting a single phenomenological parameter we can
bring GMFT into better agreement with experiments. We therefore conclude
that the long-time relaxation found in experiments can be interpreted
within GMFT as glassy dynamics consistent with the out of equilibrium
properties of a BG and its excitations.

\emph{Checkerboard case}.\textemdash The initial expansion of bosons
in the empty half of the trap in the presence of disorder is expected
to be dominated by Anderson physics, due to the low site occupations.
In order to test how enhancing interactions by increasing site occupations
affects the expansion, we have devised an ``improved'' initial state
generated by emptying sites in one half of the system according to
a checkerboard pattern. The dynamics then proceeds by allowing the
remaining bosons to evolve without any change in the Hamiltonian parameters
(no parameter quenching). Before emptying sites, the system was in
the ground state.

Fig.~\ref{fig:checkerboard} plots the normalized imbalance for the
checkerboard pattern. The pattern speeds up the decay of ${\cal I}(t)/{\cal I}(0)$
by enhancing the effect of interactions during the dynamics. It would
be interesting to find out how changes in the pattern used for the
initial state change the results in the experiments \cite{choi:2016}.

\emph{Discussion}.\textemdash Motivated by Ref.~\cite{choi:2016},
we have studied the dynamics of bosons in 2D lattices with disorder
by GMFT. We showed that theory becomes closer to experiment by including
temperature and a single phenomenological parameter. We also showed
that the features observed in the experiments are robust for various
initial states: quenched MI, disordered superfluid, and BG. Since
GMFT misses the entanglement present in MBL phases, evidence for MBL
must lie in the differences between GMFT and experiments. We find
that at the present stage with only the data from Ref.~\cite{choi:2016},
it is difficult to tell if there is a qualitative or quantitative
difference between GMFT and experiments. Further experiments, particularly
at longer times, will be needed to unambiguously show that MBL is
occurring. Avoiding macroscopic mass transport, as done in Ref.~\cite{schreiber:2015},
will help rule out slow dynamics due to Anderson and BG physics. 
\begin{acknowledgments}
M.Y., H.H., and V.W.S. acknowledge support from AFOSR (Grant No. FA9550-15-1-0445)
and ARO (Grant No. W911NF-16-1-0182), and M.R. acknowledges support
from the Office of Naval Research. We are grateful to I. Bloch, S.
Das Sarma, B. DeMarco, C. Gross, and D. Huse for discussions, and
to J. Choi for sending us details about the experimental setup \cite{choi:2016}. 

\bigskip{}
\end{acknowledgments}

\section*{Appendix: Non-quenched dynamics for uniformly distributed disorder}

\begin{figure}
\begin{centering}
\includegraphics[width=1\columnwidth]{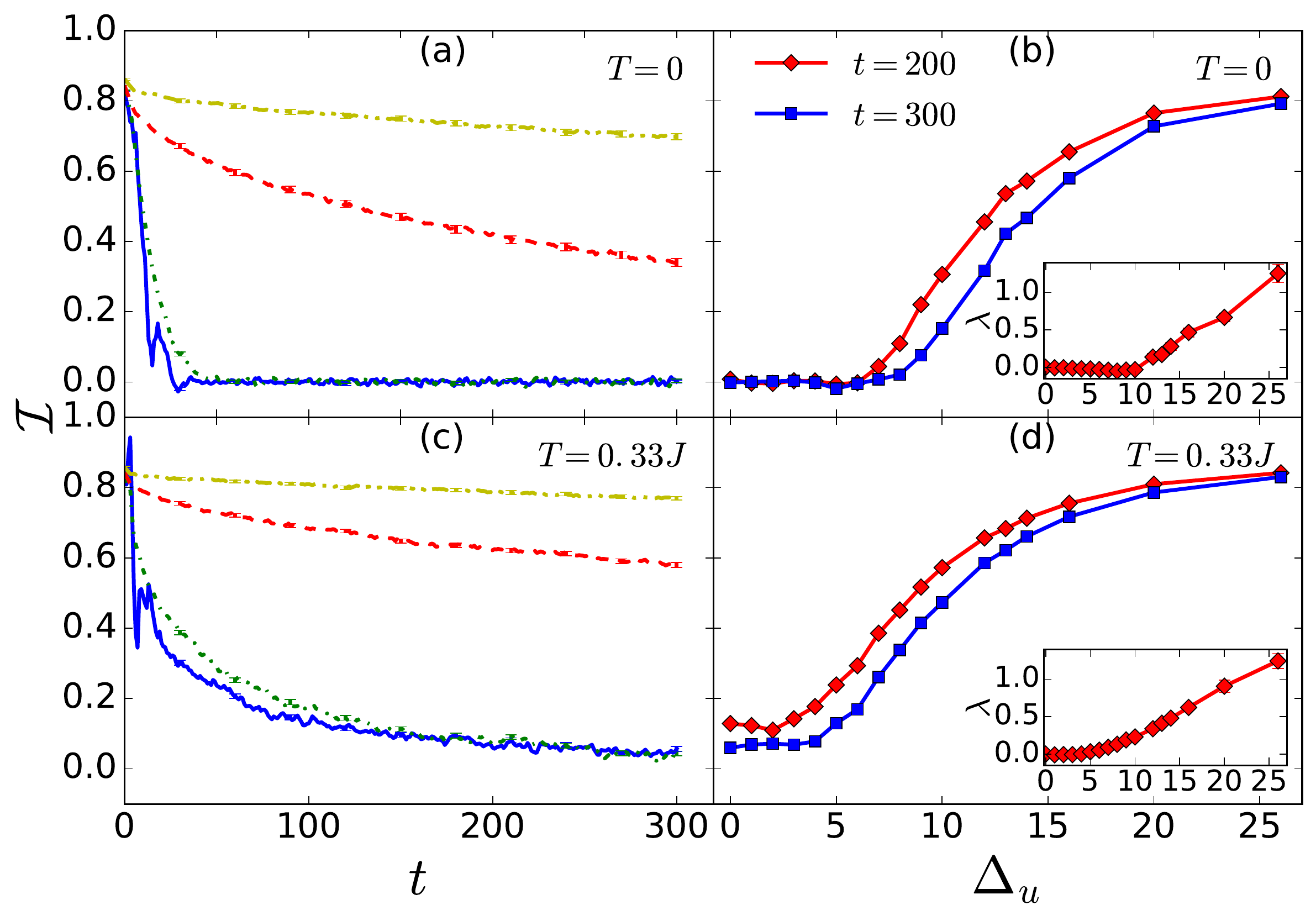} 
\par\end{centering}
\caption{\label{fig:nonquenched} Time evolution of the imbalance ${\cal I}$
for various uniformly distributed disorder strengths when the initial
temperature is $T=0$ (a) and $T=0.33$ (c). (b) and (d): Corresponding
${\cal I}$ for the same parameters but at times $t=200$ and $300$
against disorder strength. The initial state is the ground state of
Eq.(1) of the main text for $U=24.4$, $\Omega=0.145$, and $\mu=4$
in the presence of uniformly distributed disorder with strength $\Delta_{u}$.
In panels (a) and (c), the lines correspond to $\Delta_{u}=0,4,13$
and $20$ from bottom to top.}
\end{figure}

We test whether the quench protocol impacts the dynamics qualitatively.
We consider a non-quenched parameter set and study the dynamics of
the imbalance. We allow the ground state to settle into the disorder
profile before time evolving the system. The initial state is not
a Mott insulator but rather a SF (or a BG for large disorder disorder).
Here we consider the dynamics after removing all bosons in one half
of the system but without quenching any Hamiltonian parameters. For
this protocol, the initial state is selected to be the ground state
for the same values of $J$, $\Omega$, and $U$ as in the experiment
after the quench, but we trap a smaller number of bosons (the site
occupations in the center of the trap are still very close to those
in the Mott insulating state). Different from the quenched dynamics,
here we take the disorder to be distributed uniformly in the interval
$[-\Delta_{u}/2,\Delta_{u}/2]$, to show that the qualitative results
do not depend on the details of the disorder profile.

Figure~\ref{fig:nonquenched}(a) shows the time evolution of ${\cal I}$.
Comparing Fig.~2(a) of the main text and Fig.~\ref{fig:nonquenched}(a)
here one can see that the behavior of the non-quenched and quenched
cases is qualitatively similar. Quantitative differences are, on the
other hand, apparent. In the non-quenched dynamics there is no $t^{*}$
such that ${\cal I}$ does not change appreciably for $t<t^{*}$.
This is because the order parameter in the non-Mott regime is sizable.\textcolor{red}{{}
}In addition, ${\cal I}$ decays more quickly in the non-quenched
than in the quenched case. This is expected for weak disorder strengths,
for which the initial state is SF, but it is also the case in the
BG regime present for strong disorder. The results for ${\cal I}$
against $\Delta_{u}$ {[}Fig.~\ref{fig:nonquenched}(b){]} and for
$\lambda$ against $\Delta_{u}$ {[}inset in Fig.~\ref{fig:nonquenched}(b){]}
are also qualitatively similar to the corresponding graphs in Fig.~2
of the main text. The onset of the localized regime increases as $t$
increases. Figs.~\ref{fig:nonquenched}(c) and \ref{fig:nonquenched}(d)
show that the dynamics of the system slows down with the introduction
of a nonzero temperature in the initial state. This is understandable
as nonzero temperatures reduce the magnitude of the order parameter
in the SF and BG phases \cite{buonsante:2007a}. Overall, we find
no qualitative change in comparing the quenched and non-quenched cases.

\vfill{}

\bibliographystyle{apsrev4-1}
\bibliography{references}

\end{document}